\documentclass[%
 reprint,
 amsmath,amssymb,
 aps,
 pra,
]{revtex4-2}

\usepackage{graphicx}
\usepackage{dcolumn}
\usepackage{bm}
\usepackage{hyperref}

\usepackage{xcolor}
\usepackage{lipsum}
\usepackage{hyperref}
\hypersetup{
    colorlinks=true,
    linkcolor=blue,
    filecolor=magenta,      
    urlcolor=cyan,
    }
    
\begin{document}

\preprint{APS/123-QED}

\title{Photoelectron spectroscopy with a resonant dichromatic field: Role of geometric phase}

\author{Evan Munaro-Langloÿs}
 \email{evan.langloys@univ-lyon1.fr}
\affiliation{%
 Universite Claude Bernard Lyon 1, CNRS, Institut Lumière Matière, F-69622
Villeurbanne, France
}%

\author{Axel Stenquist}
\email{axel.stenquist@fysik.lu.se}
\author{Jan Marcus Dahlström}
\email{marcus.dahlstrom@fysik.lu.se}
\affiliation{
 Department of Physics, Lund University, Box 118, Lund 221 00, Sweden
}%


\begin{abstract}
We investigate geometric-phase control in resonantly driven two-level atoms exposed to near-degenerate dichromatic laser pulses. In contrast to conventional two-color schemes based on widely separated frequencies, the closely spaced frequency components generate a slowly varying beating envelope that repeatedly reverses the pseudo-spin rotation on the Bloch sphere. This enables coherent control of the geometric phase accumulated during Rabi dynamics and strongly modifies the resulting photoelectron spectra. Using an exactly solvable model for flat-top pulse envelopes, we derive the essential-state dynamics analytically and analyze photoionization induced both by an auxiliary field and by the dichromatic driving field itself. We show that the formation of Autler--Townes doublets can be interpreted in terms of destructive interference associated with geometric phases acquired during completed pseudo-spin rotations. Beyond the canonical Autler--Townes regime, beat-induced reversal dynamics lead to qualitatively different spectral structures, including re-emergent central peaks and higher-order sidebands. These effects originate from the nonuniform temporal evolution induced by the beating envelope, which modifies the balance between positive and negative excitation amplitudes. Our results establish beating-field control as a route toward engineering geometric-phase interference in ultrafast light--matter interactions and suggest broader applications in coherent control of atoms, molecules, and x-ray-driven systems.

\end{abstract}

\maketitle


\section{Introduction}
The quantum state of a qubit can be represented by a pseudo-spin vector on the Bloch sphere, $\langle\mathbf{S}(t)\rangle$ \cite{feynman_geometrical_1957}. While the expectation value of a spin transforms in the same way as a classical vector in space (SO(3) group), its quantum states generally acquire additional phase shifts.  
The quantum phase due to $360^\circ$ rotation of a spin is $\pi$, which means that the phase of the spin behaves quite like a Möbius strip (twisted belt). This SU(2) phase effect was first observed in experiments by interferometry of neutrons \cite{rauch_verification_1975}, and similar phase effects are now routinely used in quantum optics applications
\cite{nogues_seeing_1999}. The $\pi$ phase has a geometric interpretation, which in the language of Berry can be understood as due to adiabatic following of slow changes in the Hamiltonian \cite{berry_geometric_1988}. Recently, there is a rising interest to study the role of geometric phases in ultrafast processes, {\it e.g.} in high-order harmonic generation in solids \cite{uzan-narovlansky_observation_2024} and attosecond x-ray stimulated processes in molecules \cite{mi_geometric_2023}, because they are associated with robust quantum phase phenomena. 

The two-level atom subject to decay through  photoionization in intense coherent fields is a simple system to study ultrafast geometric phase effects. Here, resonant Rabi flopping leads to a $\pi$ phase for each completed Rabi cycle (full rotation of pseudo-spin vector). 
Ultrashort laser pulses in the near-infrared and optical range have allowed for numerous coherent control studies of photoionization in alkali atoms, pioneeded by Wollenhaupt and co-workers in the early 2000s \cite{wollenhaupt_quantum_2006,wollenhaupt_control_2003,bayer_bichromatic_2024}.
The Autler-Townes (AT) effect for a Rabi cycling atom was first predicted by Knight \cite{knight_saturation_1977}, and explained as due to a superposition of two dressed states \cite{knight_origin_1980}. 
From the time-domain perspective, it is known that the evolution from the single photoelectron peak to an AT doublet occurs earlier from the ground state and later from the excited state \cite{nandi_observation_2022}. At high-field intensities, the interference of photoionization by resonant and nonresonant contributions have been found to play an important role for the asymmetry of the AT peaks \cite{nandi_observation_2022,zhang_effect_2022,csehi_exact_2026} and dressed-atom stabilization \cite{beers_exact_1975,olofsson_photoelectron_2023}. Today, coherent control in noble gas atoms are attainable, following advances in intense chirped free-electron laser sources in the extreme ultraviolet range \cite{richter_strong-field_2024}, and strong-coupling phenomena have recently been reported also in the x-ray range \cite{linker_attosecond_2025}. 
Strong coupling of two-level ions following photoionization also leads to AT doublet phenomena, as predicted by Grobe and Eberly  \cite{grobe_observation_1993}. Recently, this effect has been attributed to entanglement generation between the photoelectron and the ion \cite{ishikawa_control_2023,nandi_generation_2024,jiang_time_2024,stenquist_harnessing_2025,stenquist_entanglement_2025}. Here, it has been shown that it takes only one full Rabi cycle to develop full AT peaks from the ground state, while it takes two full Rabi cycles from the excited state \cite{zhang_photoemission_2014,yu_core-resonant_2018,stenquist_harnessing_2025,stenquist_entanglement_2025}. 

In quantum optics, coherent control of qubits is central and typically requires control beyond straightforward Rabi cycling to reduce errors due to imperfections of the light and material system \cite{cruz-rodriguez_quantum_2024}. In 2019, He \textit{et al.} considered a two-level system driven by a near-resonant dichromatic laser pulse as a background-free method for observing fluorescence \cite{he_coherently_2019}. In contrast to conventional two-color schemes based on widely separated frequencies, such as fundamental and second-harmonic fields, the closely spaced frequency components considered here generate a slowly varying beating envelope. This beating leads to repeated reversals of the effective Rabi rotation on the Bloch sphere, analogous to dissonance phenomena in wave interference. The two frequencies can thereby resonantly couple to the two-level system while periodically driving the pseudo-spin vector back and forth on the Bloch sphere, as shown above panels (a) and (b) in Fig.~\ref{fig:1}. In this work, we utilize this beating-envelope dynamics to perform coherent control of the geometric phase evolution of Rabi-flopping atoms through repeated reversals of pseudo-spin rotations.


We will consider a dichromatic field in near resonance with a two-level atom and show how the amplitude and phase of photoelectrons can be controlled. 
The canonical AT doublet formation has previously been studied by an auxiliary laser field to probe its spectral build-up  \cite{jiang_enhancing_2021,zhang_monitoring_2024} and quantum phases using attosecond interferometry \cite{liao_circularly_2024}, but its geometric origin has not been explored deeply. Here, we will also apply external fields to study  photoelectron spectra affected by geometric phase beyond the canonical AT doublet.  
We find the Fourier components of the essential state dynamics for the exactly solvable case of flat-top envelopes using the Jacobi-Anger expansion, which are key to understand associated photoionization processes. 
In  Sec.~\ref{sec:Theory}, we present the theoretical background for geometric phases \ref{sec:theory-geo}, dichromatic pulses \ref{sec:theory-di-pulse}, and essential-states dynamics \ref{sec:theory-rabi-di}. In Sec.~\ref{sec:Results}, we present our results for photoionization by an auxiliary field \ref{sec:results-ft}, and intrinsic photoionization by the dichromatic field \ref{sec:results-intr}. In Sec.~\ref{sec:Discussion}, we discuss our results, and in Sec.~\ref{sec:Conclusion} we present our conclusions. Atomic units are used, $e=\hbar=m_e=4\pi\epsilon_0=1$, unless otherwise stated.

\section{Theory}\label{sec:Theory}
\subsection{Geometric phase of Rabi cycling atoms}
\label{sec:theory-geo}
It is well established that quantum state vectors differing by a complex phase 
factor represent the same physical state. A quantum trajectory $\mathcal{C}$ 
in Hilbert space satisfying $|\psi(\tau_\mathcal{C})\rangle = 
e^{i\phi}|\psi(0)\rangle$ for some time $\tau_\mathcal{C}$ therefore describes 
a system cyclically returning to the same physical state. In general, the phase 
accumulated over one cycle can be decomposed into a dynamical contribution 
$\delta$ and a geometric contribution $\beta$, as discussed by Anandan 
\cite{anandan_geometric_1992}. On one hand, the \textit{dynamical phase} is 
obtained by integrating the instantaneous expectation value of the Hamiltonian 
over one cycle,
\begin{equation}
    \delta=-\frac{1}{\hbar}\int_0^{\tau_{\cal C}} 
    \langle\psi(t)|H(t)|\psi(t)\rangle \, dt,
    \label{eq:dynamical}
\end{equation}
which reduces to the standard $e^{-iEt/\hbar}$ factor acquired by an energy 
eigenstate of energy $E$ in the case of a time-independent Hamiltonian. On the 
other hand, the \textit{geometric phase} is given by
\begin{equation}
    \beta=i\int_0^{\tau_{\cal C}}\langle \tilde{\psi}(t)|\frac{d}{dt}
    |\tilde{\psi}(t)\rangle \, dt,
\end{equation}
where $|\tilde{\psi}(t)\rangle$ is obtained from $|\psi(t)\rangle$ by an 
appropriate gauge transformation; we refer the reader to 
Ref.~\cite{Aharonov1987} for a complete derivation.

A two-level atom undergoing Rabi oscillations provides a canonical example of 
a quantum system cyclically returning to the same physical state, with the 
cycle frequency given by the Rabi frequency $\Omega$. The vector state of such a system is best described by a coherent superposition of two dressed states \cite{knight_origin_1980}, 
\begin{equation}
|\psi_\mathrm{Rabi}(t)\rangle= \sqrt{\frac{1}{2}}\left(|+\rangle e^{-i\Omega t/2}+|-\rangle e^{i\Omega t/2}\right),
\label{eq:rabi}
\end{equation}
with the initial condition $|\psi_\mathrm{Rabi}(0)\rangle=|a\rangle=(|+\rangle+|-\rangle)/\sqrt 2$, being the ground state of the atom (and $N$ photons). The excited state of the atom (with $N-1$ photons) is denoted $|b\rangle=(|+\rangle-|-\rangle)/\sqrt 2$. The dressed states $|\pm\rangle$ satisfy the eigenvalue equation, 
\begin{equation}
H|\pm\rangle=\pm\frac{\Omega}{2}|\pm\rangle,
\label{eq:eigenvalue}
\end{equation}
where $H_{aa}=H_{bb}=0$ and $H_{ab}=H_{ba}=\Omega/2$. 
The Rabi frequency is chosen as real, $\Omega=E_0 z_{ab}$, where  $z$ and $E_0$ are dipole element and electric field amplitudes, respectively. 

It is easy to show that the phases of a Rabi cycling atom are of geometric nature. We simply insert Eq.~(\ref{eq:rabi}) into Eq.~(\ref{eq:dynamical}) and use Eq.~(\ref{eq:eigenvalue}), 
\begin{equation}
    \delta_\mathrm{Rabi}= -\frac{1}{\hbar}\int_0^{\tau_{\cal C}}  \left(\frac{\Omega}{2}\langle+|+\rangle-\frac{\Omega}{2}\langle-|-\rangle\right)=0,
\end{equation}
which shows that the dynamical phase is zero. 
Thus, any net phase after a closed resonant Rabi loop has a geometric origin, $\phi_\mathrm{Rabi}=\beta_\mathrm{Rabi}$. Indeed, the phase acquired is independent of the temporal shape of the electric field amplitude (provided that the total area of the field is conserved, see Sec.~\ref{sec:theory-rabi-di}).  In contrast, an atom prepared in an eigenstate of the Rabi Hamiltonian acquires purely dynamical phase, $\phi_\pm=\delta_\pm=\mp{\Omega\tau_{\cal C}}/{2\hbar}$, with a stationary wavefunction.  In this work we will only consider resonant Rabi coupling, but it is worth to notice that if the field is detuned, then the acquired phase can have both dynamical and geometric parts.

The geometric phase for closed resonant Rabi loops follows directly from the Rabi wavefunction in Eq.~(\ref{eq:rabi})
\begin{eqnarray}
a_\mathrm{Rabi}(t)&=\langle a|\psi_\mathrm{Rabi}\rangle =   \cos(\Omega t/2),
\label{eq:rabi-amp-a}
\\
b_\mathrm{Rabi}(t)&=\langle b|\psi_\mathrm{Rabi}\rangle = -i\sin(\Omega t/2). 
\label{eq:rabi-amp-b}
\end{eqnarray}
A $N$-cyclic loop implies a time, $t=\tau_{N}=2\pi N/\Omega$, which gives the ground state amplitude, $a_\mathrm{Rabi}(\tau_{N})=\cos(\Omega \tau_{N}/2)=\exp(i \pi {N})$, in perfect agreement with the SU(2)-rotational phase. It could be noted that the phase evolution of the Rabi amplitudes are not smooth, as in the case of adiabatic following, but discrete following the sign changes of the trigonometric functions in Eqs.~(\ref{eq:rabi-amp-a}) and (\ref{eq:rabi-amp-b}), but this is no cause of alarm since the geometric phase is only defined for closed loops in the space of the physical states of the system.

\subsection{Dichromatic Pulses}
\label{sec:theory-di-pulse}

\begin{figure*}
    \centering
    \includegraphics[width=\textwidth]{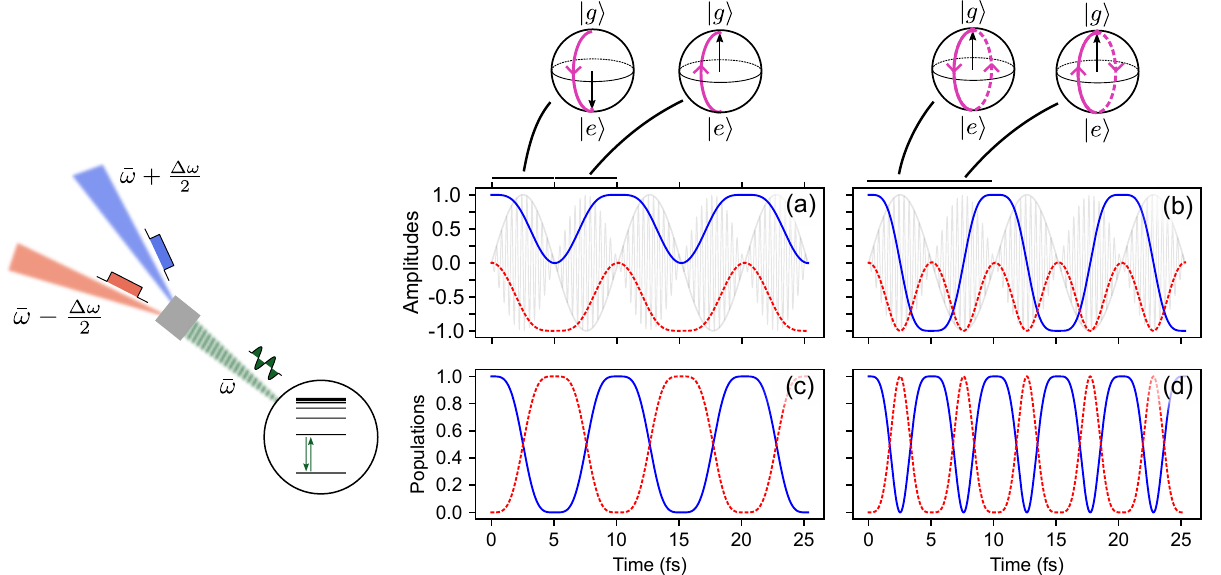}
    \caption{\textit{Typical Rabi dynamics triggered by a dichromatic pulse in a two-level atom.} The left panel shows a schematic illustration of the dichromatic field setup, where two beams of frequencies $\bar{\omega} + \Delta\omega/2$ and $\bar{\omega} - \Delta\omega/2$ are combined to produce a beating field at the mean frequency $\bar{\omega}$, which drives the transitions in a two-level atom. The two states of the two-level atom are considered to be the $1s$ and $2p$ states of hydrogen, therefore fixing $\bar\omega =$ 10.2 eV. In both cases considered here, a 25~fs pulse is used with a dichromatic parameter $\Delta\omega=$ 0.83 eV, chosen such that five full beatings occur during the pulse duration. Panels (a) and (c) show the complex amplitudes and populations, respectively, of a two-level atom interacting with a dichromatic pulse where each beating has an area of $\pi$, corresponding to a peak intensity of $3.60\times10^{13}$~W.cm$^{-2}$. Panels (b) and (d) show the complex amplitudes and populations, respectively, for a beating area of $2\pi$, corresponding to a peak intensity of $1.44\times10^{14}$~W.cm$^{-2}$. In all panels, ground state quantities are shown as solid blue lines and excited state quantities as dashed red lines. The normalized electric field envelope is displayed in the background to indicate the sign changes of the field envelope. Bloch sphere representations of the state trajectories are shown above the amplitude panels (a) and (b) for selected beatings.}
    \label{fig:1}
\end{figure*}

In this work, we define a pulse as {\it dichromatic} if it can be expressed as a superposition of two monochromatic pulses with a common envelope,  $0\le f(t)\le 1$, and distinct carrier frequencies $\omega_>$ and $\omega_<$. We define the carrier phases of the corresponding electric fields as $\phi_>$ and $\phi_<$, and set the peak amplitude of each component to $E_0/2$. In this formulation, the field $E_>(t)$ takes the form 
\begin{equation}
E_>(t) = \frac{E_0}{2} f(t) \cos({\omega_> t + \phi_>}),
\end{equation}
and similarly for $E_<(t)$. 
Their superposition yields
\begin{equation}
E(t) = E_0 f(t) \cos\left(\frac{\Delta \omega}{2} t + \frac{\Delta \phi}{2}\right) \cos({\bar{\omega} t + \bar{\phi}}) .
\end{equation}
In this expression, we define the angular frequency and phase differences as $\Delta \omega = \omega_> - \omega_<$ and $\Delta \phi = \phi_> - \phi_<$, and the corresponding averages as $\bar{\omega} = (\omega_> + \omega_<)/2$ and $\bar{\phi} = (\phi_> + \phi_<)/2$. In what follows, $\Delta\omega$ will be called the {\it dichromatic parameter}. To simplify the following discussion, we impose $\bar{\phi} = 0$ and $\Delta \phi = -\pi$, which yields
\begin{equation}
E(t) = E_0 f(t) \sin\left(\frac{\Delta \omega}{2} t\right) \cos(\bar{\omega} t).
\label{eq:electric_field}
\end{equation}
We focus on the case of a flat-top envelope of duration $T$, by defining $f(t)$ to be unity for $t \in [0,T]$ and zero elsewhere. This choice enables a fully analytical calculation of the photoionization spectrum of a two-level atom. 
We further impose $T \Delta \omega = 2\pi N_B$, where $N_B$ is an integer, to ensure continuity of the electric field---the pulse ends where the last beating ends. Clearly, $N_B$ corresponds to the number of beatings (i.e., half-oscillations of the envelope) occurring within the pulse duration. Figs.~\ref{fig:1}(a) and~\ref{fig:1}(b) show typical dichromatic pulses in gray lines, for $N_B=5$, where the optical beating arises from the superposition of two identical pulses with different carrier frequencies.

\subsection{Rabi Motion Driven by a Dichromatic Pulse}
\label{sec:theory-rabi-di}

We now consider the ``Rabi dynamics'' induced by a resonant dichromatic electric field in a two-level quantum system. To this end, we solve the time-dependent Schrödinger equation (TDSE), 
\begin{equation}
\left[i \frac{\partial}{\partial t} - \hat{H}_0 - \hat{z} E(t)\right] \psi(t) = 0.
\end{equation}
Here, $\hat{H}_0$ denotes the field-free Hamiltonian of the two-level system, with ground and excited states $|a\rangle $ and $|b\rangle$ and corresponding eigenenergies $\varepsilon_a$ and $\varepsilon_b$. The operator $\hat{z}$ represents the dipole operator, and $E(t)$ is the electric field defined in Eq.~(\ref{eq:electric_field}). We choose the dipole operator to have real matrix elements, with $z_{bb} = z_{aa} = 0$ and $z_{ab} = z_{ba}$. In the case of a dichromatic field, the resonant condition corresponds to $\bar{\omega} = \varepsilon_b - \varepsilon_a$ and the dichromatic parameter, $\Delta\omega>0$. 

Under the rotating-wave approximation, the TDSE admits the solution
\begin{equation}
\psi(t) = a(t)|a\rangle e^{-i\varepsilon_a t} + b(t)|b\rangle e^{-i\varepsilon_b t}
\end{equation}
with
\begin{equation}\label{eq:ground_state_amplitude}
a(t) =\cos\left[\frac{\Omega}{\Delta \omega}\left(1-\cos\left(\frac{\Delta\omega}{2}t\right)\right)\right]
\end{equation}
and
\begin{equation}
b(t) = -i\sin\left[\frac{\Omega}{\Delta \omega}\left(1-\cos\left(\frac{\Delta\omega}{2}t\right)\right)\right],
\label{eq:excited_state_amplitude}
\end{equation}
for a system initially prepared in the ground state. This result follows directly from the area theorem, where $\Omega = z_{ba} E_0$ denotes the Rabi frequency. The nested trigonometric structure is particularly noteworthy, as it highlights that the dynamics is governed by the dimensionless parameter $\Omega/\Delta\omega$, which controls the different dynamical regimes. Physically, this parameter represents the ratio between the characteristic timescale of the Rabi oscillations and that associated with the beating of the driving electric field. In the regime $\Omega/\Delta\omega \gg 1$, the system undergoes multiple Rabi cycles within a single beating period. Conversely, in the opposite limit $\Omega/\Delta\omega \ll 1$, the dynamics is strongly suppressed: the population transfer remains small, and the system effectively experiences only a weak, slowly varying perturbation over each beating cycle. The dichromatic field will induce slower Rabi oscillations than a flattop field of the electric field amplitude, $E_0$, due to the lower average amplitude of the effective time-dependent envelope. The dynamics are therefore characterized by the  effective Rabi frequency, 
\begin{equation}
\Omega_\mathrm{eff}= \frac{2}{\pi}\Omega, 
\label{eq:effRabi}
\end{equation}
calculated from the average absolute effective envelope (over an integer number of complete beatings, $N_B$). 

In Fig.~\ref{fig:1}, we display the complex amplitudes and corresponding populations obtained by driving a two-level system with a flat-top dichromatic pulse. The pulse parameters in Fig.~\ref{fig:1}~(a) and (c) are chosen as $\pi$-beating area---such that the resulting dynamics consists of successive half Rabi oscillations---leading to full population oscillations with no sign change of the wave function. This behavior admits a simple interpretation on the Bloch sphere: the state undergoes a back-and-forth motion along a single meridian, without completing a full rotation---no geometric phase is acquired. This geometric picture explains why the global phase of the wavefunction remains unchanged, as the trajectory never encloses a full $2\pi$ rotation and therefore does not induce a sign inversion. 

Increasing the Rabi frequency to produce $2\pi$-beating area---full Rabi oscillations during each beating---leads to amplitudes and populations shown in Fig.~\ref{fig:1}~(b) and (d), respectively. Here, it is shown that the ground state changes its sign during the evolution, while the excited state remains negative at all times. 
Thus, the Rabi frequency allows for free control of the geometric phase acquired during a beating of the dichromatic field. For the special case of an even number of beatings, the electron will always return to its initial state, leading to a closed loop in state space with zero net geometric phase. Such a pulse is an example of a zero area pulse. Even though the net geometric phase acquired after the pulse may be zero, it still plays a role during the pulse, as the electron may return multiple times to the initial state during its evolution. 
More general pulse parameters will be discussed later in the manuscript and the geometric phase will be used as an aid in our interpretation of the corresponding physical phenomena.

\begin{figure*}
    \centering
    \includegraphics[width=1\textwidth]{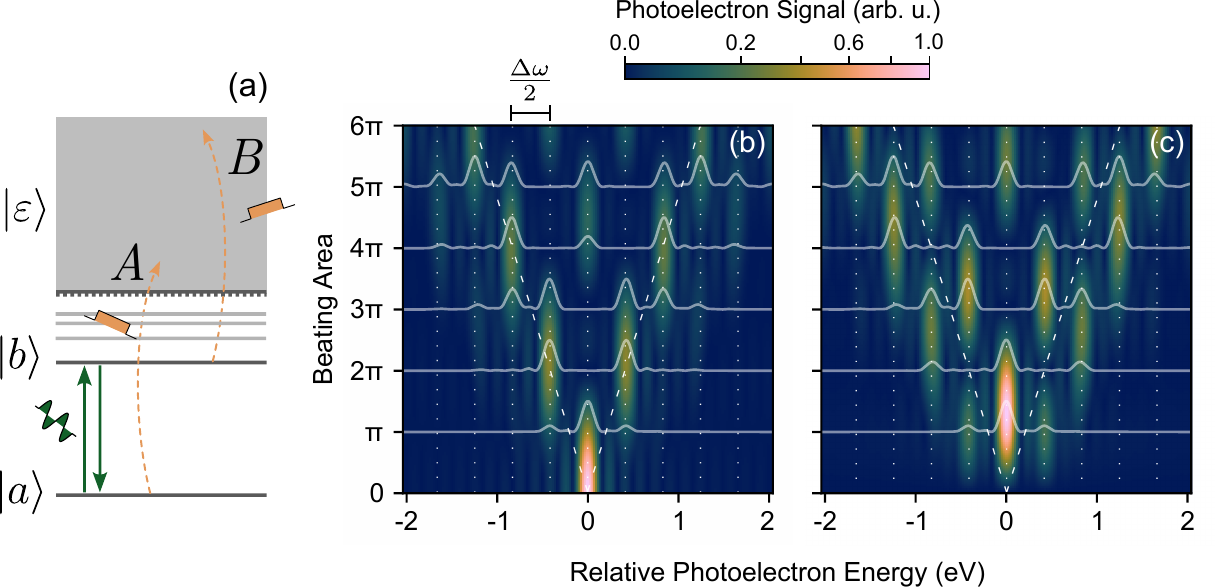}
    \caption{\textit{Photoelectron spectra induced by an auxiliary high-frequecy field, from the ground and excited states.} (a) Illustration of the ionisation by the auxiliary field (dashed arrows) from the ground and excited states $|a\rangle$ and $|b\rangle$, which are strongly coupled by the dichromatic field (solid arrows). The photoelectron spectrum corresponding to the ground and excited states are presented in panels (b) and (c), respectively. Photoelectron spectra are resolved over the pulse area of each beating.  
    In (b) and (c), normalized line shapes of the photoelectron spectra are shown at some integer multiples of $\pi$ values of the beating area. Dashed lines represent the effective Rabi frequency, $\pm\Omega_\mathrm{eff}/2$, and the dotted vertical lines are placed at half-integer and integer multiples of the dichromatic parameter, $s\Delta\omega/2$.
    }
    \label{fig:2}
\end{figure*}

\section{Results}\label{sec:Results}
Our primary objective is to assess how the interplay between the Rabi timescale and the beating timescale can be exploited to control the structure of the photoionization spectra. To this end, the average dichromatic frequency is chosen to be resonant with the transition in a two-level atom. We take model values from the 1s-2p hydrogen transition, i.e., $\bar\omega = 10.2$ eV and $z_{ba}\approx 0.745$ Bohr radius. In real atoms, the presence of additional bound states constrains the accessible dichromatic range, such that $\Delta\omega$ must remain sufficiently small to prevent the detuned component of the pulse from becoming resonant with other states, such as the $3p$ state in hydrogen. We will restrict our analysis to the two-level atom with a flat photoelectron continuum, so that the rich spectral features from the dichromatic field remain clearly identifiable. Experimentally, it is possible to increase the beating area not only by increasing the intensity of the pulse, but also by making the beating slower, by decreasing the dichromatic parameter, and the pulse longer.   

\subsection{Fourier analysis}
\label{sec:results-ft}
In this section, we study the photoelectron spectrum resulting from one-photon ionization by an auxiliary flat-top field from the ground and excited states. The two bound states are strongly coupled by the dichromatic field, but by using an auxiliary field it is possible to study ionization from each state separately \cite{zhang_monitoring_2024}, as illustrated in Fig.~\ref{fig:2}. The photon energy of the auxiliary field exceeds the system's ionization potential, such that continuum states $|\epsilon\rangle$ are reached by one photon absorption.  
Thus, the photoelectron spectrum for ionization from the ground state, $|\gamma^{(a)}(\varepsilon, t)|^2$, is attained through the Fourier transform of the ground-state complex amplitude
\begin{equation}
    \gamma^{(a)}(\varepsilon, t) \propto \int_0^{T} \mathrm{d} t' \; a(t') e^{i\varepsilon t'},
    \label{eq:photoionization_ground}
\end{equation}
where $\varepsilon$ denotes the photoelectron energy, shifted such that the center of the distribution lies at zero. The amplitude for ionization from the excited state, $\gamma^{(b)}(\varepsilon, t)$, is defined analogously. 

The integral appearing in Eq.~(\ref{eq:photoionization_ground}) cannot be evaluated analytically due to the nested trigonometric functions structure of the ground state amplitude $a(t)$, see Eq.~(\ref{eq:excited_state_amplitude}). However, it can be recast into an analytically tractable form by exploiting the Jacobi–Anger expansion,
\begin{equation}
e^{i\lambda \cos(\theta)} = \sum_{n\in\mathbb Z} i^n J_n(\lambda)e^{in\theta},
\label{eq:jacobi-anger-formula}
\end{equation}
applied to the time dependence of $a(t)$ and $b(t)$. We now detail the derivation for the ground state photoionization spectrum.
Using the trigonometric identity  $\cos(a-b)=\cos(a)\cos(b)+\sin(a)\sin(b)$, we can rewrite  $a(t)$  as
\begin{equation}
\begin{split}
a(t) = \cos&\left(\frac{\Omega}{\Delta \omega}\right)\cos\left[\frac{\Omega}{\Delta \omega}\cos\left(\frac{\Delta\omega}{2}t\right)\right] \\
+ &\sin\left(\frac{\Omega}{\Delta \omega}\right)\sin\left[\frac{\Omega}{\Delta \omega}\cos\left(\frac{\Delta\omega}{2}t\right)\right].
\end{split}
\label{eq:trig_rewrite}
\end{equation}
By identifying the real and imaginary parts of Eq.~\eqref{eq:jacobi-anger-formula}, this expression can be recast as a Fourier series:
\begin{equation}
a(t) = a_0 + 2\sum_{n=1}^{\infty} a_n \cos\left(\frac{n\Delta\omega}{2}t\right),
\label{eq:bessel_ground_state}
\end{equation}
where the coefficients  $a_n$ are time-independent and given by
\begin{equation}
a_n =
\begin{cases}
(-1)^{n/2} J_n\left(\frac{\Omega}{\Delta\omega}\right)\cos\left(\frac{\Omega}{\Delta\omega}\right), & \text{if  n  even},\\
(-1)^{(n-1)/2} J_n\left(\frac{\Omega}{\Delta\omega}\right)\sin\left(\frac{\Omega}{\Delta\omega}\right), & \text{if  n  odd}.
\end{cases}
\label{eq:coeff_a}
\end{equation}
An analogous derivation for the excited-state complex amplitude yields the expansion
\begin{equation}
i b(t) = b_0 + 2\sum_{n=1}^{\infty} b_n \cos\left(\frac{n\Delta\omega}{2}t\right),
\label{eq:bessel_excited_state}
\end{equation}
with coefficients
\begin{equation}
b_n =
\begin{cases}
(-1)^{n/2} J_n\left(\frac{\Omega}{\Delta \omega}\right)\sin\left(\frac{\Omega}{\Delta \omega}\right), & \text{if  n  even},\\
(-1)^{(n+1)/2} J_n\left(\frac{\Omega}{\Delta \omega}\right)\cos\left(\frac{\Omega}{\Delta \omega}\right), & \text{if  n  odd}.
\end{cases}
\label{eq:coeff_b}
\end{equation}
Here the time dependence is entirely contained in simple cosine functions, which can be integrated analytically. This provides direct access to a fully analytical expression for the first-order ionization amplitude in Eq.~(\ref{eq:photoionization_ground}).
Since the Bessel functions $J_n(x)$ are negligible for $|n|\gg x$ the infinite sums in Eqs.~(\ref{eq:bessel_ground_state}) and (\ref{eq:bessel_excited_state}) may be truncated.

\subsubsection{Ground-state ionization}
To attain the photoelectron spectra ionized from the ground state we insert Eq.~(\ref{eq:bessel_ground_state}) in Eq.~(\ref{eq:photoionization_ground})  
\begin{equation}
    \gamma^{(a)}(\varepsilon) = \gamma^{(a)}_0+2\sum_{n=1}^{\infty} \gamma^{(a)}_n,
    \label{eq:gamma_a_bessel}
\end{equation}
where
\begin{equation}
    \gamma^{(a)}_n = a_n\int_0^{T} \mathrm{d} t' \; \cos\left(\frac{n\Delta\omega}{2}t'\right) e^{i\varepsilon t'}.
    \label{eq:gamma_a_n}
\end{equation}
Similarly, the photoelectron spectrum corresponding to the excited state, $|\gamma^{(b)}(\varepsilon, t)|^2$, is obtained by subsituting $a$ for $b$ in Eqs.~(\ref{eq:gamma_a_bessel}) and (\ref{eq:gamma_a_n}).

The photoelectron spectrum ionized from the ground state, $|\gamma^{(a)}(\varepsilon, t)|^2$, is presented in Fig.~\ref{fig:2}(b). For $\pi$-beating area, we observe a central, resonant peak and minor symmetric side peaks.  Even though the atom is flopping between the ground and excited states several times, see Fig.~\ref{fig:1}~(c), the photoelectron spectrum does not show the canonical AT doublet. For a $2\pi$-beating area, a doublet structure is observed where the separation is the effective Rabi frequency in Eq.~(\ref{eq:effRabi}), as shown by the dashed white lines. The splitting also coincides with the dichromatic sidebands, separated by $\Delta\omega$, as shown by vertical dotted lines. The change in behaviour from a singlet to a doublet peak can be understood by analysing the ground-state amplitude. For $\pi$-beating area, see Fig.~\ref{fig:1}(a), the amplitude remains positive at all times, allowing for resonant ionization to $\varepsilon=0$. In contrast, for a $2\pi$-beating area, see Fig.~\ref{fig:1}(b), the system acquires a geometric phase---the amplitude takes both positive and negative values. This induces destructive interference at the resonant frequency of the photoelectron distribution, yielding the observed doublet structure. 
Interestingly, the effective Rabi frequency does not match with the dichromatic sidebands for odd integers of $\pi$, which leads to weaker and more scattered sideband features.
For the beating area $4\pi$ we see a re-emergence of the central photoelectron peak with side peaks at $\pm \Delta \omega\approx\pm\Omega_\mathrm{eff}/2$.
Thus, by using dichromatic pulses to strongly couple an atom, the central peak of the photoelectron distribution can be restored, a feature lacking in the canonical AT doublet. 

\subsubsection{Excited-state ionization}
The photoelectron spectrum ionized from the excited state, $|\gamma^{(b)}(\varepsilon, t)|^2$, is presented in Fig.~\ref{fig:2}(c). In this case, the spectral behaviour follows a similar trend, but is shifted to larger beating area. 
The first splitting occurs around $3\pi$-beating area. This can be understood directly from the dynamics of the excited-state amplitude $b(t)$, which does not change sign for beating areas under $2\pi$, see Fig.~\ref{fig:1}~(b). Thus, geometric phase is acquired only at larger pulse areas for the excited state, leading to destructive interference of the central peak at $\sim 3\pi$. At larger areas the central peak re-emerges, as seen at $5\pi$. We notice that the odd integer $\pi$-beating areas are the same for both the ground and the excited state due to reversal symmetry of the populations in time, see Fig.~\ref{fig:1}~(c), where an equal amount of time is spent in both states. Even integer $\pi$-beating areas do not have such symmetry, see Fig.~\ref{fig:1}~(d), where a longer time is spent in the ground- than in the excited state. Thus, different photoelectron spectra from the two states are expected, and indeed observed by comparing the corresponding line-outs in Figs.~\ref{fig:2}~(b) and (c).  

\begin{figure*}
    \centering
    \includegraphics[width=1\textwidth]{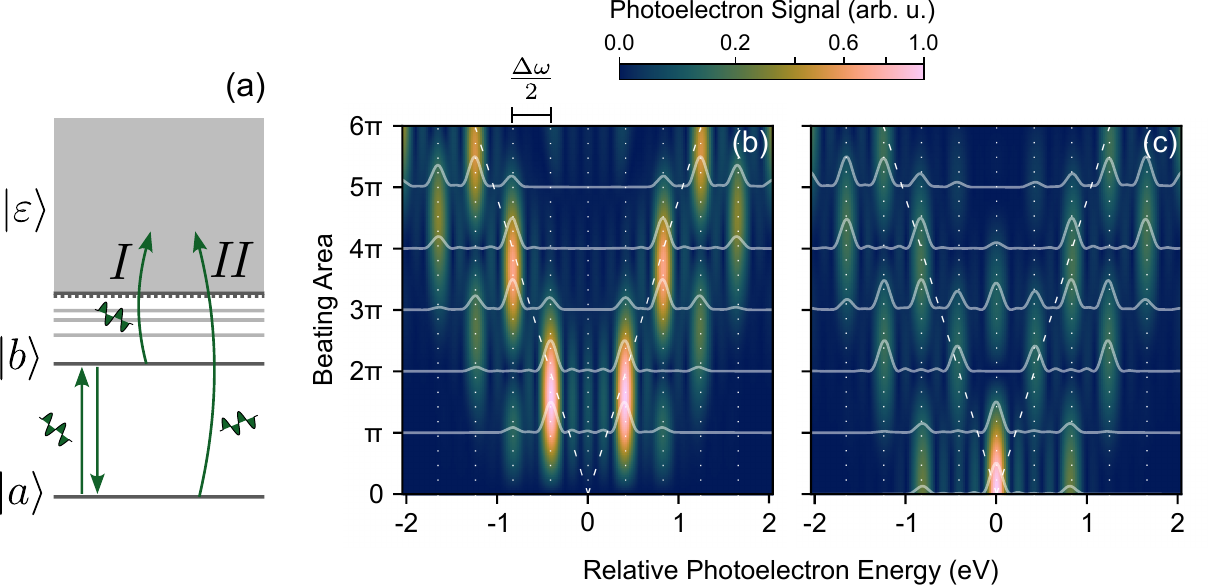}
    \caption{\textit{Photoelectron spectra generated by ionization from the dichromatic field that drives the Rabi dynamics.}
    (a) Schematic illustration of the ionization pathways and corresponding photoelectron signal. Ionization occurs from both the ground and excited states via absorption of two and one photons from the dichromatic field, respectively.
    (b) Photoelectron spectrum ionized by a one-photon transition from the excited state as a function of the pulse area of each beating.
    (c) Corresponding photoelectron spectrum from the ground state via two-photon ionization. 
    In (b) and (c), normalized line shapes of the photoelectron spectra are shown at some integer multiples of $\pi$ values of the beating area. Dashed lines represent the Rabi frequency, $\pm\Omega_\mathrm{eff}/2$, and the dotted vertical lines are placed at half-integer and integer multiples of the dichromatic parameter, $s\Delta\omega/2$.
}
    \label{fig:3}
\end{figure*}

\subsection{Intrinsic photoelectron dynamics}
\label{sec:results-intr}

To conclude this section, we investigate intrinsic ionization in a hydrogen-like two-level atom driven by the dichromatic pulse itself, as illustrated in Fig.~\ref{fig:3}(a). Ionization from the excited state proceeds via a one-photon transition, while ionization from the ground state occurs via a nonsequential two-photon process. The dichromatic nature of the ionizing field in Eq.~(\ref{eq:electric_field}) 
yields a splitting by $\pm \Delta \omega / 2$  
of the excited-state frequencies by one-photon ionization, and 
the three frequency components $-\Delta \omega$, $0$ and $\Delta \omega$ 
of the ground-state frequencies by two-photon ionization. 
To describe the ionization process, we employ time-dependent perturbation theory.  
As a zeroth-order approximation, ionization is neglected, which means that our previously obtained analytical amplitudes for the ground and excited states are valid starting points.

\subsubsection{First-order dichromatic ionization}
The one-photon ionization from the excited state is given by
\begin{equation}
\gamma^{I}(\varepsilon) \propto E_0 \int_0^T \mathrm dt' \; \sin\left(\frac{\Delta\omega}{2}t'\right) b(t')e^{i\varepsilon t'},
\label{eq:first_order_amplitude}
\end{equation}
which is the Fourier transform of the ground state amplitude split up by the dichromatic parameter, $\pm\Delta\omega/2$. 
An explicit expression for the photoionization spectra is obtained by injecting Eq.~(\ref{eq:bessel_excited_state}) into Eq.~(\ref{eq:first_order_amplitude}) yielding 
\begin{equation}
    \gamma^{I}(\varepsilon,t) = \gamma^{I}_0(\varepsilon,t) + 2 \sum_{n=1}^{+\infty}\gamma^{I}_n(\varepsilon,t),
\end{equation}
where the Jacobi-Anger terms are 
\begin{equation}
    \gamma^{I}_n(\varepsilon,t) \propto E_0b_n\int_0^t \mathrm dt' \; \sin\left(\frac{\Delta\omega}{2}t'\right) \cos\left(\frac{n\Delta\omega}{2}t'\right)e^{i\varepsilon t'},
\end{equation}
for $n=0,1,\dots$ which can be evaluated analytically.
The normalized photoelectron spectrum from one-photon ionization of the excited state, $|\gamma^I(T,\varepsilon)/E_0|^2$, is shown in Fig.~\ref{fig:3}(b) as a function of area per beating for increasing pulse intensity. For low pulse areas the spectra exhibits a doublet structure at $\pm \Delta \omega/2$ for $\pi$ and $2\pi$, and $\pm \Delta \omega$ for $3\pi$ and $4\pi$. Additionally, the spectra has integer sidebands $s$ at the energies $s\Delta\omega/2$ (vertical dotted lines). For higher areas the spectra exhibit multi-sideband structures extending outside the effective Rabi splitting (dashed lines). 

\subsubsection{Second-order dichromatic  ionization}
For the second-order dichromatic correction, two-photon ionization from the ground state is considered. Second-order ionization from the excited state is also possible, but this leads to higher energy photoelectrons via above-threshold ionization, which is beyond the scope of the present analytical work. After adiabatic elimination of the intermediate states, {\it c.f.} Refs.~\cite{Kaufman2020,zhang_effect_2022}, the second-order amplitude reads
\begin{equation}
    \gamma^{II}(\varepsilon) \propto E_0^2 \int_0^T \mathrm dt' \; \sin^2\left(\frac{\Delta\omega}{2}t'\right) a(t')e^{i\varepsilon t'},
    \label{eq:second_order_amplitude}
\end{equation}
which depends on the square of the beating envelope, {\it i.e.} addition of three shifted Fourier transforms of the ground state amplitude. As for the first order, injection of Eq.~(\ref{eq:bessel_ground_state}) into Eq.~(\ref{eq:second_order_amplitude}) gives an explicit expression for the ionization amplitude:
\begin{equation}
    \gamma^{II}(\varepsilon,t) = \gamma^{II}_0(\varepsilon,t) + 2 \sum_{n=1}^{+\infty}\gamma^{II}_n(\varepsilon,t)
\end{equation}
where the Jacobi-Anger terms are pairs of triplets 
\begin{equation}
    \gamma^{II}_n(\varepsilon) \propto E_0^2a_n\int_0^T \mathrm dt' \; \sin^2\left(\frac{\Delta\omega}{2}t'\right) \cos\left(\frac{n\Delta\omega}{2}t'\right)e^{i\varepsilon t'}.
\end{equation}
for $n=0,1,\dots$ which can be evaluated analytically.

The spectrum for two-photon ionization from the ground state, $|\gamma^{II}(T,\varepsilon)/E_0^2|^2$, is presented in Fig.~\ref{fig:3}(c). Contrasting behaviour to the one-photon ionization is observed. At small beating area, the spectrum exhibits a clear triplet structure. This structure is most simply explained as due to the combination of the frequency components of the two photons involved in the transition from the unperturbed ground state. Similarly, the one-photon dichromatic doublet at low areas in Fig.~\ref{fig:3}~(b) is simply due to the two colours of the dichromatic field from the unperturbed excited state. 

\begin{figure}
    \centering
    \includegraphics[width=1\linewidth]{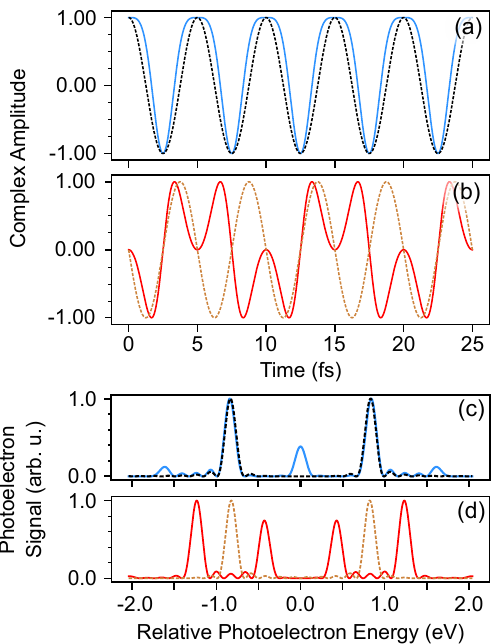}
    \caption{\textit{Comparison of Rabi dynamics induced by the dichromatic field and a corresponding flat-top field.} Panels (a) and (b) show the time evolution of the ground and excited state complex amplitudes, respectively, when driven by a 25~fs dichromatic pulse with five beatings and a beating area of $4\pi$. The dashed lines show the corresponding complex amplitudes when driven by a flat-top pulse designed such that the Rabi frequency matches $\Omega_\mathrm{eff}$. Panels (c) and (d) show the photoelectron spectra corresponding to ionization from the ground and excited states, respectively, by an external probe field, analogous to the case presented in Fig.~\ref{fig:2}. Solid and dashed lines correspond to photoelectron spectra obtained from a system driven by the dichromatic pulse and the flat-top pulse, respectively.}
    \label{fig:4}
\end{figure}

\section{Discussions}\label{sec:Discussion} 
In Fig.~\ref{fig:2} we observed that the photoelectron spectra splits into a doublet around $2\pi$-beating area for electrons ionized by an auxiliary field from the ground state (b) and around $3\pi$ from the excited state (c). This $\pi$-area delay originates from the initial transfer dynamics to reach the excited state. The splitting can be understood in the time domain as destructive interference due to the geometric phase induced by a full pseudo-spin-rotation. 
We also noticed that beyond this formation of doublet structures, the central peak re-emerges at $4\pi$ for the ground state and $5\pi$ from the excited state. Although a $4\pi$ pulse area corresponds to two completed Rabi cycles, the geometric phase is playing a role. The resulting photoelectron spectrum differs qualitatively from the canonical AT doublet. 
Amplitude dynamics are presented in Fig~\ref{fig:4}(a) for $4\pi$ beating area (full line). The dynamics are compared with ordinary Rabi oscillation with the effective Rabi frequency (dashed line). 
In ordinary Rabi oscillations the amplitude spends roughly the same amount of time in the positive, $a(t)>0$, and negative, $a(t)<0$, regime, yielding destructive interference of the central peak.
However, due to the sinusoidal behaviour of the dichromatic envelope the dynamics slow down when the rotation changes direction, close to $a(t)\approx 1$ in this case. This implies that the system spends a longer time in the positive-amplitude regime, which allow for the formation of a central peak.
In Fig.~\ref{fig:4}(b) the complex amplitude of the excited state deviates strongly from the a ordinary Rabi amplitude. Due to symmetry the amplitude spends equal amount of time being positive and negative, yielding destructive interference of the central peak. In Fig.~\ref{fig:4}(c) and (d) we present the photoelectron distribution ionized with an auxiliary field from ground- and excited state, respectively. The dichromatic field with $4\pi$-beating area (full line) is compared with the AT doublet induced by ordinary Rabi oscillations (dashed line). As predicted, the ground state has a central peak in the dichromatic case, which is not observed for ordinary Rabi oscillations. The remaining spectra matches the AT doublet structure (the ordinary Rabi frequency was chosen to match the effective Rabi frequency of the dichromatic pulse). As predicted, the excited state has no central peak, but exhibits four sidebands---showing strong contrast with the usual AT doublet.  

\section{Conclusion}\label{sec:Conclusion} 
We have found that the geometric phase can be used to interpret complex photoelectron spectra from resonantly coupled atoms driven by dichromatic pulses. The beating of the effective envelope leads to a lower effective coupling strength, but it also provides rich interference patterns that provide spectral information about the electron dynamics. The dimensionless ratio of the Rabi period and the beating period was found to be a useful quantity to study coherent control of a two-level atom and its associated photoelectron spectra. Interestingly, the photoelectron spectra differ quantitatively for ionization by an auxiliary field and by the dichromatic field itself. In the future, application of dichromatic pulses for coherent control may find other applications in more complex systems, such as vibrating molecules, because it can be used to remove geometric phase effects induced by the laser coupling and reduce spectral congestion. Another natural continuation of this work would be to consider non-resonant dichromatic pulses to study photoelectron spectra influenced by a combination of geometric and dynamical phases.

\textit{Acknowledgements}---JMD acknowledges support from the Knut and Alice Wallenberg Foundation: 2024.0212 and the Swedish Research Council: 2024-04247. EML thanks Victor Despré and Franck Lépine for helpful discussions and support.

\bibliography{Dichromatic}

\end{document}